\documentclass[11pt,letterpaper]{elsarticle}

\usepackage[
  top=1in,
  bottom=1in,
  left=1in,
  right=1in
]{geometry}

\usepackage{amssymb}
\usepackage{float}
\restylefloat{table}
\usepackage[utf8]{inputenc}
\usepackage[T1]{fontenc}
\usepackage{textcomp}
\usepackage{soul}
\usepackage{graphicx,subfig}
\usepackage{placeins}
\usepackage{amsmath,array,amssymb,xparse,upgreek,bm}
\usepackage{tikz,mathpazo}
\usetikzlibrary{shapes.geometric,arrows}
\usepackage{hyperref}
\usepackage{tcolorbox}
\usepackage{booktabs}
\usepackage{multirow}
\usepackage{upgreek}

\usepackage[ruled]{algorithm2e}
\usepackage{empheq}
\usepackage{tablefootnote}
\usepackage{array}

\journal{JOSS draft}

\begin{document}

\begin{frontmatter}

\title{IMRFit: Interactive Simulation and Constitutive-Model Fitting for Inertial Microcavitation Rheometry}

\author[1]{Sicong Wang}
\author[1]{Zixiang Tong}
\author[1,2]{Jin Yang}

\address[1]{Department of Aerospace Engineering and Engineering Mechanics, University of Texas at Austin, Austin, TX, USA}
\address[2]{Texas Materials Institute, University of Texas at Austin, Austin, TX, USA}


\end{frontmatter}

\section{Summary}

Inertial microcavitation rheometry (IMR) characterizes the high-rate mechanical response of soft materials from the rapid expansion and collapse of a microscopic bubble \cite{gaudron2015bubble,estrada_high_2018,yang_extracting_2020,tzoumaka2023modeling}. In a typical IMR experiment, high-speed imaging provides the bubble radius as a function of time, $R(t)$, and material parameters are identified by comparing the measured response with simulations that couple bubble dynamics to a constitutive description of the surrounding medium. This inverse analysis requires repeated forward simulations and careful control of experimental inputs, material parameters, and fitting settings.

\textbf{IMRFit} (https://github.com/YangMechanicsGroupUTAustin/IMRFit) is an open-source Python desktop application that brings these steps into an interactive environment for forward simulation, parameter identification, and constitutive-model comparison. Users can import experimental radius--time data, select a bubble-dynamics formulation and constitutive model, inspect simulated responses, fit material parameters, and organize batch fitting and parameter-sweep jobs. IMRFit includes Neo-Hookean Kelvin--Voigt~\cite{estrada_high_2018} and generalized Maxwell--Ogden-based models~\cite{wang2026experimental}, among other constitutive options, and supports additional models through JSON descriptors and Python solver entry points. It also integrates fitting-window selection, visualization, and data export into a single workflow.

\begin{figure}[h!]
\centering
\includegraphics[width=1 \textwidth]{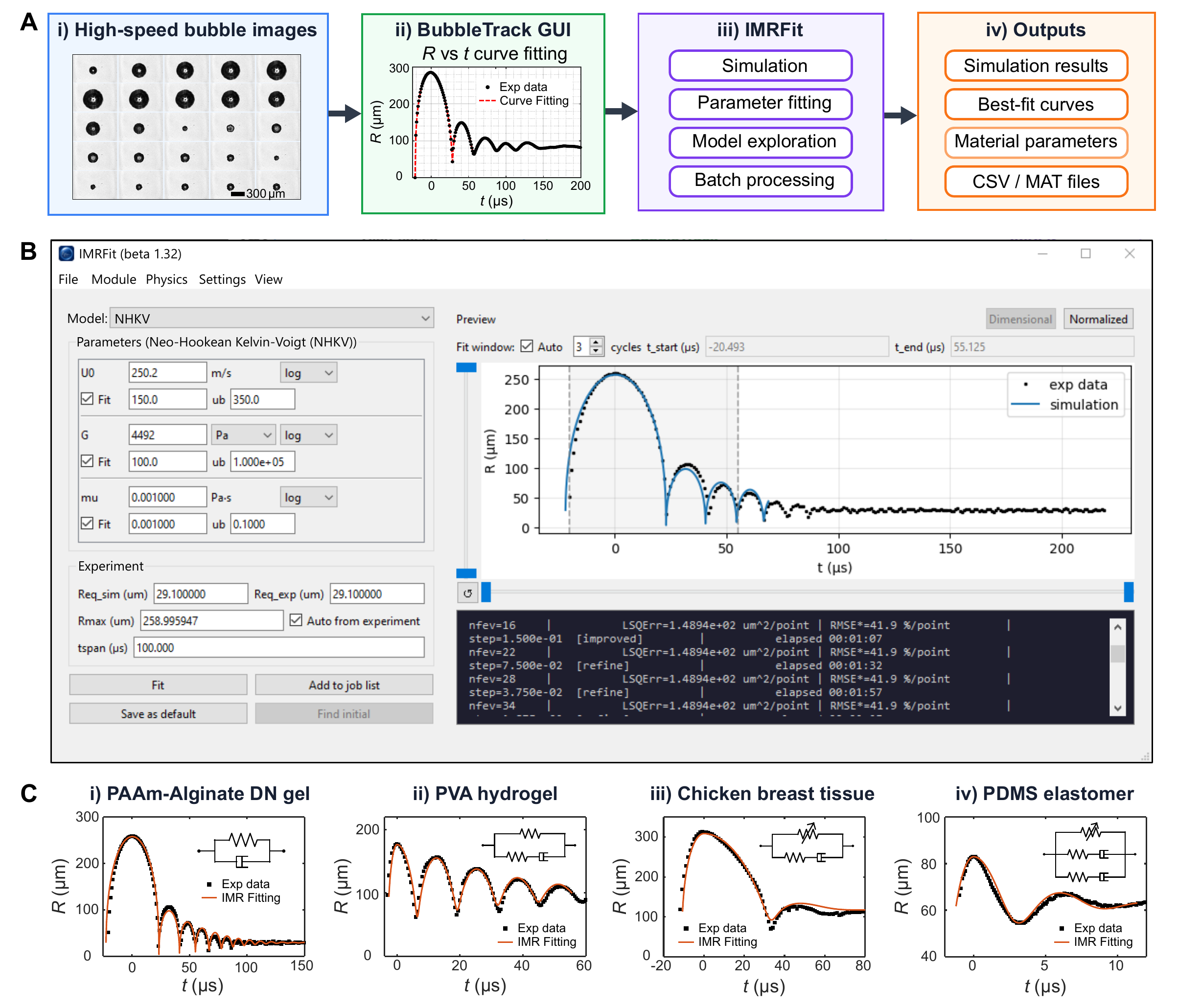} 
\caption{\textbf{IMRFit workflow, graphical interface, and representative applications.} (A) End-to-end IMR analysis: (i) high-speed imaging of bubble expansion and collapse; (ii) extraction of radius--time data using BubbleTrack; (iii) forward simulation, parameter fitting, constitutive-model exploration, and batch processing using IMRFit; and (iv) export of simulated curves, fitted parameters, and analysis results. (B) IMRFit graphical interface showing model and parameter selection, experimental and simulated radius--time curves, fitting-window controls, and optimization progress. (C) Representative fits to experimental measurements (black markers) using IMRFit simulations (red curves): (i) polyacrylamide (12 wt\%)--alginate (0.3 wt\%) double-network hydrogel with the Neo-Hookean Kelvin--Voigt (NHKV) model~\cite{bu2025bubble}; (ii) poly(vinyl alcohol) (PVA) hydrogel~\cite{wang2026experimental} and (iii) chicken breast tissue with a one-branch generalized Maxwell--Ogden damage (GMOD) model~\cite{wang2026experimental}; and (iv) polydimethylsiloxane (PDMS) elastomer with a three-branch standard nonlinear Ogden damage (SNOD) model. Insets show the corresponding constitutive-model architectures.}
\label{fig:overview}
\end{figure}

As illustrated in Fig.~\ref{fig:overview}A, IMRFit complements \textbf{BubbleTrack}, a separate application for extracting calibrated bubble radius--time curves from high-speed image sequences \cite{bubbletrack}. BubbleTrack converts images into experimental measurements, while IMRFit uses those measurements for simulation and material-parameter identification. Together, the two applications connect image-based measurement to mechanical inference while retaining a modular interface between the stages.

\section{Statement of need}

Identifying material properties from IMR measurements is a nonlinear inverse problem~\cite{estrada_high_2018,sanchez2026hierarchical}. For each trial parameter set, the bubble-dynamics equations must be solved and the predicted radius--time curve compared with experimental measurements over a selected fitting interval. Beyond the numerical optimization itself, the inferred parameters depend on choices such as unit conversion, equilibrium-radius and acquisition metadata, initial guesses, parameter bounds, optimizer settings, and the fitting window. In script-based workflows, these choices may be distributed across multiple files, making repeated analyses and comparisons difficult to manage consistently.

Moreover, a small numerical fitting error does not necessarily imply that the selected constitutive model adequately captures every feature of the experiment. Collapse timing, rebound amplitude, oscillation damping, and relaxation toward equilibrium may respond differently to changes in model parameters. Researchers therefore need to examine candidate fits, assess parameter sensitivity, and determine where a model agrees or disagrees with the measured response. These tasks benefit from an environment in which experimental curves, model assumptions, parameter settings, and numerical predictions can be inspected together.

IMRFit addresses these practical needs for researchers and students in experimental mechanics, soft-material rheology, and constitutive-model identification. By placing routine operations in a consistent graphical workflow, it reduces the need to modify analysis scripts for individual datasets and facilitates systematic comparisons across experiments and models.

\section{State of the field}

IMR modeling and analysis have advanced alongside experimental studies of transient cavitation in soft materials \cite{estrada_high_2018,yang_extracting_2020,tzoumaka2023modeling,luo2020laser,estrada2021neural,yang2022mechanical,buyukozturk2022particle,yang2024estimating,zhu2025parsimonious,chu2026limits,sreedasyam2026scanning,yang2021predicting,mcghee2023high,yang2022probing,kaushik2024radial,abeid2024ultra,yang2025inertial,abeid2024experimental,mancia2020amr,spratt2021characterizing,wang2026laser,yang2019semannual,yang2020semannual,kolluri2025determination,remillard2026microbubble,remillard2026nonspherical,chu2026accelerating,chu2025bayesian,bremer2026repetitive}. Existing software includes the MATLAB-based IMRv2 project~\cite{imrv2} for IMR simulation and analysis and the Python-based \textbf{PyIMR} package, which emphasizes differentiable forward models, sensitivities, and inference-oriented computation~\cite{bryngelson2026pyimr}. These resources provide important numerical capabilities but are organized around different user workflows.

IMRFit is designed specifically for interactive, experiment-centered analysis. Its contributions include guided experimental-data import, direct control of fitting windows and constitutive parameters, reusable fitting queues, and side-by-side comparison of measured and simulated curves. Rather than introducing a competing bubble-dynamics solver, the application integrates numerical functionality available through SciPy \cite{virtanen_scipy_2020} with Matplotlib visualization \cite{hunter_matplotlib_2007} and domain-specific data management. Its distinguishing contribution is therefore the organization of IMR fitting and model exploration into an accessible, extensible desktop workflow.

\section{Software design}

IMRFit separates experimental-data handling, constitutive-model definitions, numerical simulation, and fitting-job management. Its desktop interface is built with PySide6 and embedded Matplotlib plots. The main workspace (Fig.~\ref{fig:overview}B) displays the selected constitutive model and adjustable parameters alongside experimental and simulated radius--time curves, fitting-window controls, and optimization progress. This arrangement allows users to evaluate the influence of parameter choices and model assumptions without switching between scripts and separate plotting tools.

The Import Wizard reads MATLAB MAT files with varying field names and supports data recorded in either SI or micrometer--microsecond units. It can apply pixel-to-length calibration, reconstruct the time axis from frame-rate information when necessary, and perform optional data cleanup. Curves can be displayed in dimensional or normalized form, allowing experimental measurements and simulations to be compared consistently.

The numerical layer supports Rayleigh--Plesset and Keller--Miksis bubble-dynamics formulations \cite{rayleigh_viii_1917,plesset_dynamics_1949,keller_bubble_1980} coupled to constitutive models for the surrounding medium. Built-in options include the Neo-Hookean Kelvin--Voigt (NHKV), standard nonlinear solid (SNS), generalized Maxwell--Ogden (GMOD), and standard nonlinear solid 3-branch (SNS3) and standard nonlinear Ogden damage (SNOD) models. These formulations enable investigation of nonlinear elasticity, viscoelasticity, and damage-like softening; the implemented viscous branches use linear dashpots. Models are registered through JSON descriptors that define their parameters, default values, bounds, display units, scaling rules, and Python solver entry points. This separation enables researchers to add constitutive models without modifying the main graphical interface.

Parameter identification minimizes a least-squares mismatch between measured and simulated radii within the specified fitting interval. Interactive and queued fits use the same objective-function logic, so their results can be compared on a consistent basis. Users may define fitting windows manually or select a prescribed number of bubble oscillation cycles. The job manager retains the experimental data, model identity, parameter values and bounds, physical constants, optimizer settings, fitting interval, and output metadata for each job. It also supports batch fitting, parameter sweeps, job reloading, and queue archiving. A separate Curve View panel facilitates comparison of multiple curves and export to MAT and SVG formats.

\section{Research impact statement}

IMRFit has been used to fit radius--time measurements from soft materials exhibiting substantially different bubble responses. Figure~\ref{fig:overview}C(i)--(iv) shows representative comparisons for a polyacrylamide--alginate double-network hydrogel fitted with NHKV, a PVA hydrogel and chicken breast tissue fitted with a one-branch GMOD model, and a PDMS elastomer fitted with a three-branch SNOD model. These examples illustrate the use of a common interface to analyze different experiments and evaluate constitutive assumptions of varying complexity. They demonstrate software functionality; agreement between an individual fitted curve and the data does not, by itself, establish parameter identifiability or the uniqueness of a constitutive model.

The principal research contribution of IMRFit is a reusable environment for organizing and evaluating IMR inverse analyses, rather than a new bubble-dynamics equation or constitutive theory. By keeping model settings, fitting choices, simulated responses, and exported results connected within one application, it supports repeatable analyses and more transparent comparisons between candidate models. Its extensible model interface also provides a practical framework for evaluating additional constitutive descriptions against experimental bubble dynamics.

\section{AI usage disclosure}

IMRFit was developed primarily through iterative, prompt-directed collaboration between S.W. and OpenAI Codex (GPT-5.5 for the initial application framework and most of the implementation, and GPT-6-sol for later revisions). S.W. specified the scientific models, software requirements, and user workflows; Codex generated and revised application code, tests, and documentation in response. S.W. reviewed the changes and tested the software and numerical results.

BubbleTrack, the companion application that provides radius--time data to IMRFit, was developed by Z.T. with assistance from Claude Code (Anthropic), using the Claude Opus 5.5 model. Claude Code was used to generate graphical-interface components and to implement and refactor image-processing routines for bubble segmentation and radius extraction. Z.T. designed the software architecture and algorithms and reviewed and tested AI-generated code. 

OpenAI ChatGPT (GPT-6) assisted with language editing of this manuscript. The authors reviewed and edited AI-assisted material and take responsibility for the software, scientific models, numerical verification, interpretation of results, and final manuscript content.

\section*{Acknowledgements}

J.Y. gratefully acknowledges support from the U.S. National Science Foundation under Grant Nos. CMMI 2232428 and 2441460; the U.S. Office of Naval Research under Grant No. N000142612190, through Dr. Timothy Bentley; and the U.S. Army Research Office under Award No. W911NF-26-1-A274, through Dr. Denise Ford.

\bibliographystyle{unsrt}
\bibliography{reference}

\end{document}